\documentclass{aa}  

\usepackage{graphicx}
\usepackage{txfonts}
\usepackage{lipsum}
\usepackage{xcolor}
\usepackage[hidelinks,colorlinks=true,linkcolor=blue,citecolor=blue]{hyperref}
\usepackage{subcaption}         
\usepackage{lscape}             
\usepackage{placeins}           
\usepackage{ulem}  
\usepackage{amsmath}

\begin{document}
   \title{Old pulsar wind nebulae and the role of the thermal filaments}


   \author{N. Bucciantini\inst{1,2,3}\fnmsep\thanks{email: niccolo.bucciantini@inaf.it}
        \and Y. Batini\inst{2,1} 
        \and B. Olmi\inst{1}
        }

\institute{
INAF, Osservatorio Astrofisico di Arcetri, Largo E. Fermi 5, I-50125 Firenze, Italy
\and
Dipartimento di Fisica e Astronomia, Universit\`a di Firenze, Largo E. Fermi 2, I-50125 Firenze, Italy
\and
INFN , Sezione di Firenze, Via G. Sansone 1, I-50019 Sesto Fiorentino (FI), Italy
}

   \date{Received XX XX, 20XX}

 
  \abstract
   {Old pulsar wind nebulae are among the foremost galactic high-energy gamma-ray sources. However we still lack a robust and reliable approach to their modeling, especially in the light of forthcoming high-energy observatories like Astri Mini-Array or CTAO. Part of the problem is due to the complex interaction that characterizes these systems. Understanding this complexity has then become mandatory for further advancements. }
   {We aim to develop a new approach to investigate the possible role that the thermal thick layer of massive filaments, seen in objects like the Crab nebula and 3C 58, but likely present in all pulsar wind nebulae, can exert on the dynamics of the late reverberation phase, and compare results with standard approaches that neglect the presence of such layer.} 
   {A new formulation of the one-zone thin-shell  plus Lagrangian formalism, that we developed in a series of previous papers, is here extended to the case of a thick layer of filamentary ejecta, complementing our former work, that was mostly focused on the initial free-expansion phase.}
 {We compare the dynamics of reverberation with and without the presence of filaments, and show that in the former case, not only reverberation may be substantially anticipated ($\sim 30\%$), but that the following compression takes a much longer time. On the other hand the total compression of the system does not seem to change much, and the qualitative behavior is preserved. }
   {Our results suggest that the presence or absence of
   an extended filamentary layer might affect the duration of both the free-expansion phase (shortening it) and that of the following compression phase during reverberation (lengthening it), but it does not change much the overall compression of the nebula. While this changes the relative number of systems in these two phases, and their 
   contribution to high-energy emission, some peculiar radiative effects associated to the level of compression in old systems, like the "super-efficiency" might not be much affected.}

   \keywords{ pulsar: general -- method: numerical -- ISM: supernova remnants }

   \maketitle
\nolinenumbers

\section{Introduction}
\label{sec:intro}
Pulsar Wind Nebulae (PWNe) are bubbles of relativistic pair plasma, that originate from the confinement of the pulsar (PSR) wind by the parent supernova remnant (SNR) of its stellar progenitor \citep{Gaensler_Slane06a}. 
While young systems are mostly observed as powerful non-thermal synchrotron sources, typically in radio and X-rays \citep{Kargaltsev15}, older systems are mostly revealed in the TeV gamma-ray domain. This is where the  relic low energy leptons accumulated over the life of the system \citep{Olmi23a} produce radiation through inverse Compton scattering on seed photons, most frequently photons from the Cosmic Microwave Background (CMB) or thermal infrared ones (both in the far and near infrared, FIR/NIR).
Given the estimated rate of core-collapse supernovae \citep{Rozwadowska_Vissani+21a}, we expect roughly 2000 potential gamma-ray emitting systems in the Galaxy \citep{Fiori_Olmi+22a,DeSarkar26a}, of which we currently detect and identify around thirty \citep{Kargaltsev_Rangelov+13a}. Moreover, the population of ultra-high energy gamma-ray sources (with emission above 100 TeV up to PeVs) recently detected by LHAASO approaches a hundred to date, with the indication that many of them are directly connected with pulsars \citep{LHAASOCat:2024}. 
These numbers are bound  to increase substantially in the upcoming future with the operation of new TeV observatories as the Cherenkov Telescope Array Observatory \citep[CTAO][]{CTA+19a} or the ASTRI Mini-Array \citep{Astri22+}, as well as the continuation of LHAASO operations. 
A correct modeling of the expected emission from these systems is then mandatory, both to assess the chance of future detection, as well as to  properly characterize any observable target.\\
\\
Current strategies for the modeling of old systems mostly rely on the so called one-zone thin-shell approach \citep{Ostriker_Gunn71a,Reynolds_Chevalier84a,Gelfand_Slane+09a,Bucciantini_Arons+11a,Martin_Torres+12a}. While this approach has proved quite robust for young systems, its reliability for older ones has always been an iffy business, with different groups resorting to different recipes, leading often to contrasting results \citep{Gelfand_Slane+09a,Bucciantini_Arons+11a,Martin_Torres+12a}. Much of this stems from the fact that while for young systems the PWN and the SNR are causally disconnected, and can be evolved separately, at later times their evolution becomes coupled, and their mutual feed-back is highly non-linear. Even if one neglects the pulsar proper motion, or gradients in the density of the ambient medium, there is no simple analytical way of parameterizing this interaction. \\
\\
Simple 2D, and even 3D, hydrodynamical simulations of old PWNe have been presented in the past \citep{Blondin_Chevalier+01a,van_der_Swaluw-Downes+04a,Kolb_Blondin+17a,Meyer_Meliani+24a,Meyer_Torres25a}, but no attempt to use them to derive their expected spectral properties has been done, mostly because, in order to make such simulations numerically manageable, several strong simplifications have to be done (adiabaticity, no magnetic field, non-relativistic fluids, no pulsar spin-down), to the point that they can only be relied on at a qualitative level, but cannot be used to make robust quantitative predictions. \\
\\
The standard one-zone thin-shell models that are used to describe the evolutions of PWNe inside SNRs, are based on the assumption that during the initial free-expansion phase, all of the SNR ejecta swept-up by the PWN accumulate in a thin shell at the very boundary of the nebula itself \citep{Ostriker_Gunn71a,Reynolds_Chevalier84a,Gelfand_Slane+09a,Bucciantini_Arons+11a,Martin_Torres+12a}. Recently, in a series of papers \citep{Bandiera_Bucciantini+20a,Bandiera_Bucciantini+23a,Bandiera_Bucciantini+23b} it has been shown that, in the following reverberation phase, this shell roughly maintains its integrity, at least as long as the dynamics is one-dimensional. In those same studies it was shown that the shell does not accumulate matter during the reverberation phase, and \textit{de-facto}, it can be treated as a separate structure with respect to both the PWN on the inside and the SNR shell on the outside. These findings inspired the development of a new strategy to model the dynamical and spectral evolution of PWNe inside SNRs: the PWN is treated using the standard one-zone approach, with or without radiation losses, as a piston that acts on a thin-shell of swept-up ejecta; the SNR is instead evolved using the equation of ideal hydrodynamics, and subject to inner boundary conditions dictated by the dynamics of the thin-shell (this might be easily done adopting a Lagrangian formalism);  the thin-shell itself evolves according to the mass and momentum conservation laws, under the action of the PWN on its inner side and the SNR on its outer side. The set of equations forms a closed system, that can be evolved in time \citep{Bandiera_Bucciantini+23b,Bucciantini_Olmi26a,DeSarkar26a}.\\
\\
However, in a recent paper \citep[][Paper I hereafter]{,Bucciantini_Olmi26a}, this entire picture has been put into question, in relation to the early free-expansion phase, and possibly to the later reverberation phase too. It has been known, for a long time indeed, that in the Crab nebula (and 3C~58 too) the swept-up ejecta are not confined in a thin shell at the boundary of the relativistic PWN, but form a layer of thermally emitting filaments/knots that penetrate deeply inside the PWN \citep{Trimble68a,Hestewr-Strone+99a,Fesen_Rudie+08a,Owen_Barlow15A,Martin_Milisavljevic+21a,Martin_Milisavljevic+25a}. Moreover, recent mass estimates seem to suggest that most of the swept-up mass is in these deeply embedded filaments and not in the outer skin \citep{Owen_Barlow15A}. 
Driven by these and other considerations, we have developed a new formalism that allows one to salvage the one-zone approach, with its versatility in terms of spectral evolution, and still model the development of a thick layer of mixed ejecta \citep{Hestewr-Strone+99a,Jun98a,Bucciantini_Amato+04a,Porth_Komissarov+14a}. 
Application of the new formalism to the Crab nebula  suggested that, once the presence of a layer of filaments was accounted for, the evolution of a PWN in the free-expansion could be quantitatively different from the one predicted using the standard thin-shell approach. That work, however, left unanswered the question on how relevant the presence of a thick layer, in place of a thin-shell, could be for the following reverberation phase.\\
\\
Unfortunately, while there is some evidence for mixing also in evolved system \citep{Ma_Ng+10a}, foremost the large extension observed in a few TeV PWNe \citep{HESS19a,Aharonian06a,HESS19b}, there is really no system that can offer us some guidance on how to modify our theoretical approach, on par with the Crab nebula. Moreover due to the large numerical cost of simulating old PWNe, even numerical results \citep[e.g.][]{Blondin_Chevalier+01a,Kolb_Blondin+17a,Meyer_Torres25a} do not provide us a robust and reliable picture of their evolution. 
In the light of this uncertainty, we present here a modified version of the standard one-zone thin-shell approach to the reverberation phase that accounts for the fact that when the PWN reaches the SNR reverse shock, there is no massive shell but an extended layer of filaments.\\
\\
The paper is organized as follows: In Sect.~\ref{sec:model} the equations for the model are introduced; in Sect.~\ref{sec:sefsim} we present and discuss our results for a selected sample of models spanning the expected PWN-SNR population, and compare them with the canonical thin-shell results; finally in Sect.~\ref{sec:conclusions} we summarize our conclusions.
\section{The model}
\label{sec:model}
\begin{figure}
\includegraphics[width=9cm]{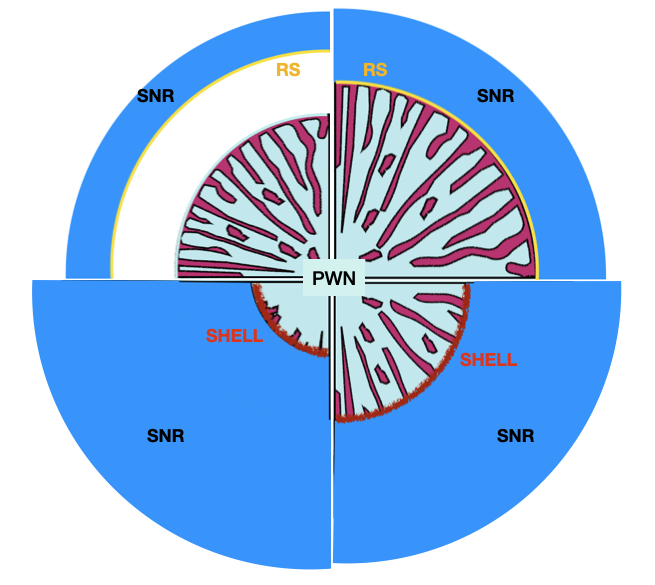}
    \caption{Schematic evolution of a PWN inside a SNR in the presence of an extended layer of high density filaments that penetrate the PWN itself. Top-left panel: initial structure in the free expansion phase, an SNR shell is present bounded on the inside by a reverse shock (RS), while a high density PWN with extended high density fingers is expanding outward. Top-right panel: the moment when reverberation begins, and the PWN reaches the reverse shock. Bottom-right panel: compression of the PWN by the SNR following reverberation: the reverse shock has disappeared, and now matter in the filaments accumulates in a thin shell at the boundary between PWN and SNR. Bottom-left panel: the compression of the PWN has reached its maximum, and almost all matter in the filaments has now been accumulated in the thin shell, { even if some filaments might still be present in the central region (not shown in figure)}.}
    \label{fig:schema}
\end{figure}

We model the PWN-SNR system with three fluids (see Fig.~\ref{fig:schema} for a schematic representation of the structure of this system during its evolution), with different levels of coupling among them: 
\begin{itemize}
    \item The PWN is treated as a uniform bubble containing a relativistically hot massless plasma, whose energy varies due to adiabatic losses (and radiation too) and injection by the PSR. 
    During the initial free-expansion phase the PWN is not
    well coupled to the surrounding cold ejecta, that instead of accumulating in a thin shell at the PWN boundary, end up forming a layer of filaments, that penetrate it, and over which it exerts a reduced drag. {  If the swept up mass ends up in fingers that penetrate inside, and that expand at a lower speed than the PWN edge \citep{Martin_Milisavljevic+25a}, then the kinetic energy of this material is lower than in the canonical model, where it accumulates all in a thin shell at the edge. This means that the energy transferred by the PWN is also reduced, implying a lower coupling between the PWN and the cold SNR ejecta.}
    During the later reverberation phase on the other hand,  the PWN is coupled
    with those ejecta that have been shock heated by the SNR reverse shock. The reason for this difference can be attributed to the different magnetization of the latter to the former. While it is expected that the cold ejecta that give rise to the thermal filaments should be almost completely unmagnetized, and  this holds for many of the high density thermal knots \citep{Hestewr-Strone+99a,Owen_Barlow15A}, there are some evidence that the plasma downstream of the reverse shock is magnetized \citep{Gotthelf_Koralesky+01a,DeLaney_Koralesky+02a,Rho_Dyer+02a}, which will ensure a much better coupling with the non-thermal PWN. This means that a well defined boundary will from between the PWN and the reverse shock hearted ejecta, and this boundary, in the late reverberation phase, can be taken as the radius of the PWN, $R_{\rm pwn}$.
    \item The SNR shell is treated as an ideal classical fluid, obeying the equations of 1D radial fluid dynamics. The matter in the SNR shell, especially the one downstream of the reverse shock, is coupled to both the PWN and the filaments.
    It is unclear the level of coupling of the two,
    and it is possible that the highest density knots, { if produced by Rayleigh-Taylor instability from the interaction of the PWN and SNR,} might freely penetrate the SNR in the form of shrapnels \citep{Park_Hughes04a,DeLaney_Rudnick10a,Miceli_Orlando13a}. For simplicity we assume that, having similar densities, the filaments do not efficiently penetrate the SNR shell, and at most accumulate at its inner boundary $R_{\rm in}$, which is also the edge of the PWN.
    \item The filaments are treated as a cold pressureless dense plasma, characterized by a high compressibility that, while uncoupled to the PWN, interact efficiently with the SNR shell  accumulating in a thin shell at a radius $R_{\rm sh}$, coincident with the SNR inner boundary. { We assume that the filaments penetrate all the way to the center. In reality, the fingers might only partially penetrate the PWN, as in the case of the Crab nebula where they are seen to extend from 0.3-0.4$R_{\rm pwn}$ outward \citep{Martin_Milisavljevic+21a}. However, as shown in Paper I, the presence of such a small central region, free of filaments, does not change significantly the overall dynamics. In this sense, our assumption amounts to maximizing the mixing, and can be seen as the opposite limit with respect to the canonical thin shell model, where the swept-up matter accumulates all at the edge. }
\end{itemize}
The tree components are then all coupled together at:  $R_{\rm pwn} = R_{\rm in} = R_{\rm sh}$. There a shell will form. The equations describing the evolution of this shell are the same used in the standard thin-shell formalism. The mass in the shell will grow due to accumulation of the matter from the filaments:
\begin{equation}
  \frac{{\rm d}}{{\rm dt}}  M_{\rm sh}(t) = 4\pi R_{\rm sh}(t)^2  v_{\rm rel} (t) \,\rho_{\rm fil}(R_{\rm sh},t)\;\Theta[v_{\rm rel}(t)],
  \label{eq:mass1}
\end{equation}
where $v_{\rm rel}(t) = v_{\rm fil}(R_{\rm sh},t)-\dot{R}_{\rm sh}(t)$ is the relative velocity between the shell and the filaments, whose velocity profile is given by $v_{\rm fil}(r,t)$, and whose density profile is $\rho_{\rm fil}(r,t)$.  
On the other hand, the shell momentum evolves due to momentum transfer from  the accumulating filaments, and under the action of the PWN pressure on the inside and the SNR pressure on the outside:
\begin{align}
     \frac{{\rm d}}{{\rm dt}}   [M_{\rm sh}(t) \dot{R}_{\rm sh}(t)] &= 4\pi [P_{\rm pwn}(t)-P_{\rm snr}(t)] R_{\rm sh}(t)^2  +\nonumber\\
     &\quad +\frac{{\rm d} M_{\rm sh}(t)}{{\rm dt}}   v_{\rm fil}(R_{\rm sh},t).
\end{align}
If one assumes a fully developed  homogeneously expanding filamentary layer of uniform  density, as the one considered in Paper I, then one can introduce the auxiliary radius: 
\begin{equation}
    \tilde{R}(t) = R_{\rm pwn}(t_{\rm beg,rev})+\dot{R}_{\rm pwn}(t_{\rm beg,rev}) (t-t_{\rm beg,rev})\;,
\end{equation}
where $t_{\rm beg,rev}$ is the time at which reverberation begins, defined by the condition $R_{\rm pwn}(t) = R_{\rm rs}(t)$, with $R_{\rm rs}(t)$ the reverse shock radius. 
{$\tilde{R}$ is the maximum radius that the filaments (equal to the radius of the PWN) would have had if the PWN had not collided with the reverse shock, but had instead moved outward at the velocity it had at the moment of collision with the reverse shock.}
As in Paper I, here we  have assumed that, during the free-expansion phase ($t<t_{\rm beg,rev}$), the outer edge of the filaments corresponds to the outer edge of the PWN, and { that it moves} 
at the same speed. 
Then:
\begin{equation}
    v_{\rm fil}(R_{\rm sh},t) = \dot{R}_{\rm pwn}(t_{\rm beg,rev}) \frac{R_{\rm sh}(t)}{\tilde{R}(t)},
\end{equation}
and if one assumes a standard  core-envelope profile for the SNR ejecta (again as was done in Paper I), { Eq.~\ref{eq:mass1} can be solved and} one finds:
\begin{equation}
    M_{\rm sh}(t) = \tilde{M}_{\rm ej}\left(\frac{R_{\rm pwn}(t_{\rm beg,rev})}{v_{\rm c} t_{\rm beg,rev}} \right)^{3-\delta}\left[1-\left(\frac{R_{\rm sh}}{\tilde{R}(t)} \right)^3\right],
\end{equation}
where $v_{\rm c}$ is the velocity at the { boundary of the core region of the ejecta, enclosing a mass $\tilde{M}_{\rm ej}$. These equations can be generalized to more complex distributions of the filaments, including the case of partial penetration.}
\\
To close this system one needs an equation for the PWN pressure, and in the non radiative case one can compute it according to:
\begin{equation}
    4\pi \frac{{\rm d}}{{\rm dt}}   [R_{\rm sh}^3(t)P_{\rm pwn}(t)] = L_{\rm psr}(t) - 4\pi P_{\rm pwn}(t) R_{\rm sh}(t)^2 \dot{R}_{\rm sh}(t).
\end{equation}
{ The SNR pressure just outside the PWN, $P_{\rm snr}(t)$ is derived by solving the full 1D evolution of the SNR, for which we employ the Lagrangian scheme described in \citet{Bandiera_Bucciantini+23a}.}

\section{Comparison with the canonical thin-shell approach}
\label{sec:sefsim}
\subsection {Population synthesis}
\begin{figure}
\includegraphics[width=9cm]{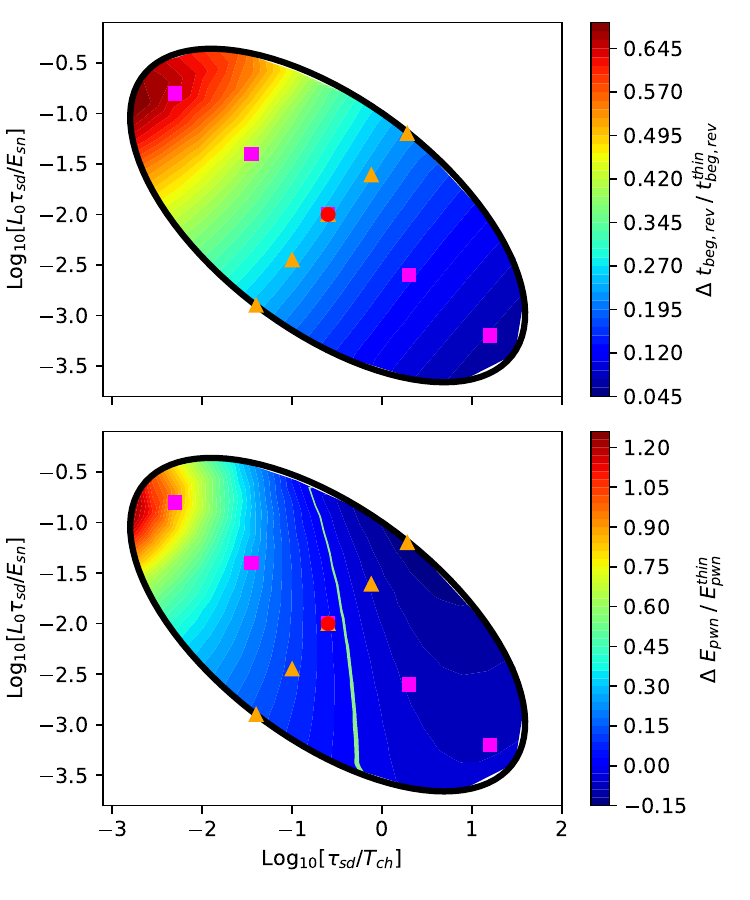}
    \caption{Comparison of canonical thin-shell models and thick-shell ones ($X=0.6$) for the PWN-SNR population. The black ellipses represents the region in the characteristic plane containing 98\% of the expected Galactic PWN-SNR population. Points represent the models of Tab.~\ref{table:1}: the red dot at the center is \texttt{C0}, yellow tringles represent models of the sequence  \texttt{S\#}, and magenta squares those of \texttt{L\#}. The upper panel represents the relative difference in the time at which reverberation begins $\Delta t_{\rm beg,rev} = t_{\rm beg,rev}^{\rm thin}-t_{\rm beg,rev}^{\rm thick}$. The lower panel represents the relative difference in the energy content of the PWN at the time at which reverberation begins $\Delta E_{\rm pwn} = E_{\rm pwn}^{\rm thick}-E_{\rm pwn}^{\rm thin}$. The green line separates systems with positive difference from systems with negative difference. }
    \label{fig:popul}
\end{figure}

In order to make a meaningful comparison between the canonical approach to the reverberation and our new strategy, we have first computed a representative population of PWN-SNR systems. This is done following recipes similar to the ones present in the literature  \citep{Fiori_Olmi+22a,Bandiera_Bucciantini+23a, DeSarkar26a}, but without accounting for their distribution within our Galaxy, their magnetization or any other input quantity related to their emissivity, given that we are not interested here in computing their observability, and we will operate in the non-radiative regime. In this case the only parameters of interest are the pulsar initial spin-down power $L_0$, the spin-down time $\tau_{\rm sd}$, the SNR energy $E_{\rm sn}$, the ejecta mass $M_{\rm ej}$ and the ambient medium density $\rho_{\rm ism}$ (which can be used to define a characteristic time $T_{\rm ch}$ and radius $R_{\rm ch}$, \citealt{Truelove_McKee99a}).  
The pulsar spin-down properties are given by its initial period and magnetic field, whose distributions are taken from \citet{Watters_Romani11a}. For the initial period we set a lower limit of 10~ms, as in the original work by \citet{Watters_Romani11a}, while we set an upper limit of 500~ms, above which there are mostly magnetars \citep{Mereghetti_Pons+15a}. For the magnetic field, the lower limit was set at $10^{11}$~G, compatible with the minimal values of observed PSRs within PWNe (lower values are more characteristic of thermal neutron stars that show no evidence of pair creation activity, \citealt{Pavlov_Sanwal04a,Ho_Wynn13a}), and an upper limit at $4.4\times 10^{13}$~G given that pair creation above the QED critical field is suppressed by photon splitting \citep{Timokhin_Harding15a,Timokhin_Harding19a}. 
For the spin-down law we assume a standard dipole braking index, $n=3$. 
Moreover, we adopted a standard PSR radius of $10$~km \citep{Ozel_Freire16a},  a typical PSR mass $M_{\rm psr} = 1.40$~M$_\odot$ \citep{Biwas_Rosswog25a}, and a moment of inertia $I=10^{45}$~g~cm$^{2}$ \citep{Breu_Rezzolla16a}. The SNR energy is taken from a uniform distribution in the range $(0.2-2.0) \times 10^{51}$~erg~s$^{-1}$ \citep{Sukhbold_Ertl+16a}. The ejecta mass is equal to $M_{\rm fin} - M_{\rm psr}$ where, in order to account for mass losses, the final mass of the progenitor star $M_{\rm fin}$ is a function of the zero age main sequence mass $M_{\rm zams}$ according to \citet{Sukhbold_Ertl+16a}, and the latter is sampled from a Salpeter initial mass function \citep{Salpeter55a} in the range $9-25$~M$_\odot$ \citep{Sukhbold_Ertl+16a}. Given the large uncertainty in pre-supernova mass loss rates, and more recent estimates that suggest reduced losses due to wind clumpiness \citep{Renzo_Ott+17a}, we also considered the case of no mass loss at all ($M_{\rm fin}=M_{\rm zams}$), and found that the distribution of the final population was not altered in any significant way. The ejecta density follows a core-envelope profile, as in Paper I, with the inner core slope $\delta =0$ and envelope slope $\omega =12$ \citep{Matzner_McKee99a}. For the ISM density instead we used the numerical results by \citet{Kim_Ostriker17a} for the local Galactic medium. We assumed that one third of supernovae are, in the nomenclature of \citet{Kim_Ostriker17a}, "runaway" (i.e. they sample the ISM in an unbiased way), and the rest classifies as "cluster" \citep{Chu_Gruendl08a,Carretero-Castrillo_Ribot+24a}. Again we compared this choice with the simple analytical prescription by \citet{Elwood_Murphy+17a}, based on  SNR modeling, and found no significant difference in the distribution of the PWN-SNR population.
Our synthetic population allows us to define a region of interest in the characteristic plane $L_0 \tau_{\rm sd}/E_{\rm sn}$ - $\tau_{\rm sd}/T_{\rm ch}$ \citep{Bandiera_Bucciantini+23a}, that will contain 98\% of the expected Galactic PWN-SNR systems, as shown in Fig.~\ref{fig:popul}.\\
\\
The upper-left part of this region represents very energetic PSRs, that moreover release all of their energy on timescales much smaller than the SNR characteristic time. They behave in an \textit{explosive} fashion, and their energetics is strong enough to affect the global SNR dynamics. 
However, we do not expect more than a few systems like these to have formed in the past 100~kyr in the Galaxy. This means that it is unlikely that we will be able to see any of them early enough to appreciate this explosive interaction. On the other hand, systems at the bottom-right of the population, have PSRs that not only inject little energy, but do so on very long timescales. For these we expect the SNR to evolve almost unaffected by the PWN.  \\
\\

\subsection{Conditions at the beginning of the reverberation}
In Fig.~\ref{fig:popul} we plot, in this region of interest, both the difference in the time it takes the system to reach reverberation, $t_{\rm beg,rev}$, and the difference in the pressure of the PWN at the beginning of reverberation between the canonical thin-shell case, and the case of a fully developed homologously expanding filamentary layer (corresponding to $X=0.6$ in the equations for the thick-shell of Paper I).
This gives an overview of the differences already present at the beginning of the reverberation, for systems characterized by the exact same parameters, due just to different recipes for the behavior of the swept-up ejecta.

In the presence of a thick filamentary layer, the PWN reaches the reverse shock earlier. In the case of very energetic systems, this can be about 50\% earlier than in the thin-shell case, and typically happens when the reverse shock is still moving outward, while for low energy systems the difference is at most 10\%, and reverberation begins when the reverse shock is moving inward. The difference in $t_{\rm beg,rev}$ implies a difference in the size of the PWN at reverberation, $R_{\rm beg,rev}$. It is also evident that for the time difference there is a clear trend along the major axis of our elliptical region. However this difference is not that pronounced, because the dynamical trajectory of the reverse shock, whose radius reaches a maximum and then begins to recede, changes by less than 20\% between 0.5 and 1.75 $T_{\rm ch}$. For the vast majority of systems the maximum PWN size is unaffected by the presence or not of a thick-filamentary layer. In Tab.~\ref{table:1} a comparison for equivalent models in the two regimes is provided, for the two sequences of Fig.~\ref{fig:popul}, tracing systems along the major and minor axis of the region of interest. One sees that the size of the PWN at $t_{\rm beg,rev}$ does not change significantly along the region minor axis, while some stronger variations are seen along the major one, with models at the beginning and end of the sequence reverberating with a smaller PWN radius, either because they encounter the reverse shock very early or very late. In Fig.~\ref{fig:popul}, we also show the difference in the energy content of the PWN. Despite the fact that in the presence of a thick layer the PWN reaches reverberation earlier, its energy content can be almost twice larger for energetic systems. For low energy systems instead, the trend is reversed. However differences are less than 15\% for the vast majority of systems, and almost zero at the peak of the distribution (the center of the region).

\subsection{Evolution during reverberation}

\begin{figure}
\includegraphics[width=9cm, bb=1 40 420 200, clip]{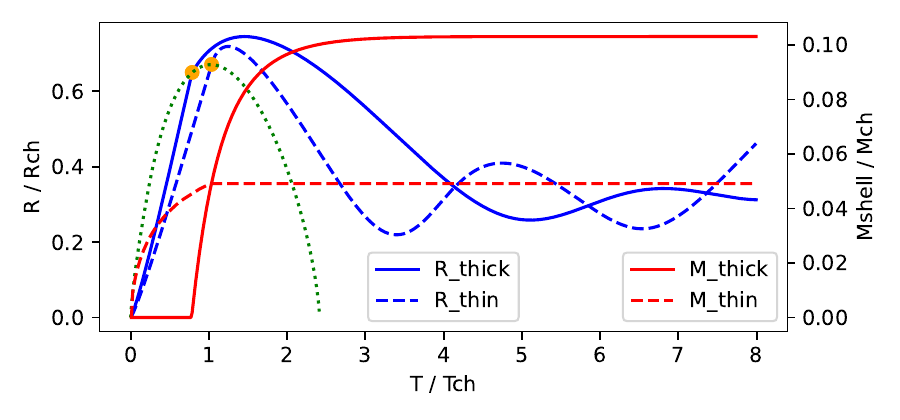}
\includegraphics[width=9cm, bb=1 40 420 200, clip]{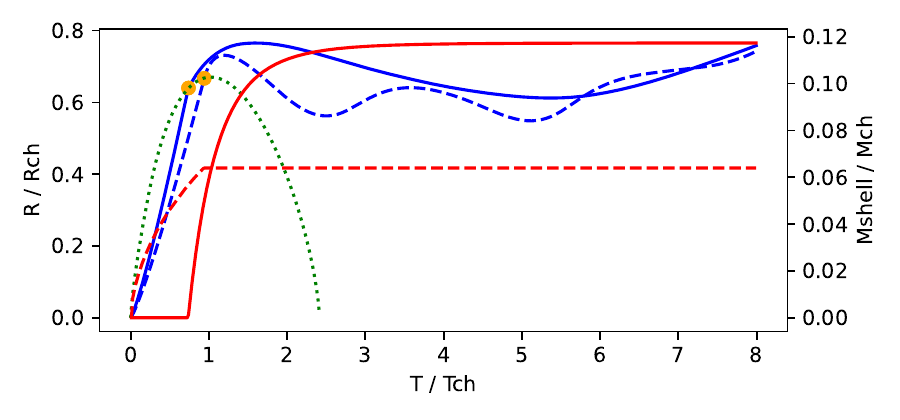}
\includegraphics[width=9cm, bb=1 10 420 200, clip]{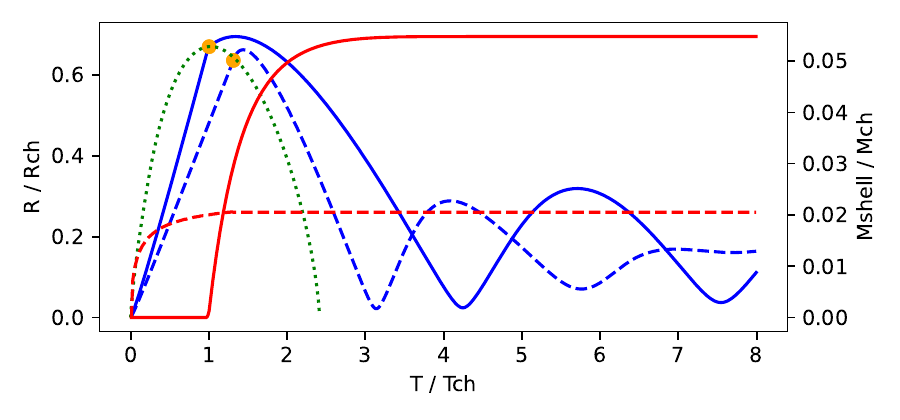}
    \caption{Evolution of the PWN radius (blue) and mass on the outer PWN shell (red), for both the standard one-zone thin-shell formalism dashed), and in the case of a thick homogeneous fully developed filamentary layer (solid).  The yellow dots represent the moment when reverberation begins $t_{\rm beg,rev}$. From top to bottom: models \texttt{C0,S1,S4}. The dotted (green) curve describe the trajectory of the reverse shock for an SNR with $\delta=0$ and $\omega=12$. The characteristic mass is by definition the mass $M_{\rm ch}$ of the SNR ejecta, while the characteristic radius $R_{\rm ch}$ can be derived from the SNR energy, ejecta mass and ISM density \citep{Bandiera_Bucciantini+21a}.}
    \label{fig:evol}
\end{figure}

In Fig.~\ref{fig:evol} we show the evolution of the radius of the PWN for models \texttt{C0, S1, S4}, along the minor axis, that span a wide range of compressions from just 30\% to about a factor 30, as shown in Tab.~\ref{table:2}. A similar trend holds also along the major axis from model  \texttt{L4} to model \texttt{L1}, and similar differences between the canonical thin-shell and our thick-shell formalism are present. 
Focusing on the model \texttt{C0} we see that in the thick case, reverberation is not only anticipated, but the PWN tends to expand to a slightly larger radius, and this is reached at a later time, compared to the maximum of the thin case. 
Interestingly, the evolution of the radius in the thick case, appears to be a delayed and stretched version of the thin one. The first compression takes place at $\sim 5T_{\rm ch}$, is followed by a re-expansion and then a second compression sets in. The series of compressions and re-expansions is very similar to what is typically found for the equivalent thin-shell model, and the radii of the first minima are roughly the same. There is however some indication that the amplitude of these oscillations is reduced. This is possibly due to the fact that unlike the thin-shell case, where the mass of the shell is saturated at the beginning of the reverberation, for the thick-layer case, the mass of the shell at the PWN outer surface grows in time, probably leading a more relaxed behavior. Interestingly, for a thick-layer case it takes no more than one characteristic time for the mass accumulated at the outer PWN boundary to almost saturate, and this mass reaches values that are about twice what is found in the thin-shell case. \\
\\
This overall qualitative trend holds also for less and more compressive systems. For model \texttt{S1}, where compression during reverberation is marginal, the dynamics of the PWN radius in the presence of a thick-layer does not show the even minor repeated compressions and re-expansions of the thin-shell case, and seems to relax quite smoothly to the long term trend. 
For the strongly compressive \texttt{S4} model the radial evolution follows very closely, even if in a somewhat delayed fashion, the thin-shell case. 
Interestingly, the mass accumulated at the PWN edge in all cases is roughly twice the mass of the thin-shell. The fact that both the maximum and minimum radius of the PWN are not much affected by the way the swept up mass behaves, implies that the evolution of the PWN pressure (and hence the magnetic energy), is also quite similar. And indeed we found that the maximum pressure during compression changes less than a factor 2.

\begin{table*}[ht!]
\caption{Models and comparison of the parameters at the beginning of the reverberation: "thin" and "thick" refers  to the standard thin-shell formalism and the one presented here for a thick-layer respectively.}                 
\label{table:1}    
\centering                        
\begin{tabular}{c c c c c c c }      
\hline\hline \\[-1.7ex]           
Model&  $Log_{10}(\tau_{sd}/T_{\rm ch})$ & $Log_{10}(L_0\tau_{\rm sd}/E_{\rm sn})$ &  $t^{\rm thin}_{\rm beg,rev}/T_{\rm ch}$ & $R^{\rm thin}_{\rm beg,rev}/R_{\rm ch}$ &  $t^{\rm thick}_{\rm beg,rev}/T_{\rm ch}$ & $R^{\rm thick}_{\rm beg,rev}/R_{\rm ch}$\\         
\hline \\[-1.7ex]                        
\texttt{C0} & -0.60 & $-2.00$ & 1.033 & 0.671 & 0.785 & 0.650\\  
\texttt{S1} &  ~0.28 & $-1.20$ & 0.940 & 0.667 & 0.739 & 0.640\\ 
\texttt{S2} & -0.12 & $-1.60$ & 0.987 & 0.670 & 0.770 & 0.647\\ 
\texttt{S3} & -1.00 & $-2.45$ & 1.157 & 0.660 & 0.878 & 0.663\\ 
\texttt{S4} & -1.40 & $-2.90$ & 1.312 & 0.636 & 1.002 & 0.670\\ 
\texttt{L1} &  ~1.20 & $-3.20$ & 1.870 & 0.409 & 1.793 & 0.491\\ 
\texttt{L2} &  ~0.30 & $-2.60$ & 1.498 & 0.596 & 1.281 & 0.640\\ 
\texttt{L3} & -1.45 & $-1.40$ & 0.661 & 0.625 & 0.398 & 0.495\\ 
\texttt{L4} & -2.30 & $-0.80$ & 0.351 & 0.456 & 0.134 & 0.233\\ 
\hline  \vspace{2pt}   
\end{tabular} 
\end{table*}

\begin{table*}[ht!]
\caption{Models and comparison of the properties of compression during reverberation: "thin" and "thick" refers respectively to the standard thin-shell formalism, and the one presented here for a thick-layer; "max" and "min" refers to the conditions when the PWN reaches its maximum size and its first minimum after compression. The compression factors $CF$ are defined as the ratio of the minimum over maximum radius. }                 
\label{table:2}    
\centering                        
\begin{tabular}{c c c c c c c c c c c}      
\hline\hline \\[-1.7ex]           
Model&  $t^{\rm thin}_{\rm max}/T_{\rm ch}$ & $R^{\rm thin}_{\rm max}/R_{\rm ch}$ &  $t^{\rm thick}_{\rm max}/T_{\rm ch}$ & $R^{\rm thick}_{\rm max}/R_{\rm ch}$ & $t^{\rm thin}_{\rm min}/T_{\rm ch}$ & $R^{\rm thin}_{\rm min}/R_{\rm ch}$ &  $t^{\rm thick}_{\rm min}/T_{\rm ch}$ & $R^{\rm thick}_{\rm min}/R_{\rm ch}$ & $CF^{\rm thin}$ & $CF^{\rm thick}$\\         
\hline \\[-1.7ex]                        
\texttt{C0} & 1.250 & 0.719 & 1.452 & 0.745 & 3.405 & 0.219 & 5.111 & 0.258 & 3.278 & 2.885\\  
\texttt{S1} & 1.204 & 0.731 & 1.607 & 0.766 & 2.490 & 0.562 & 5.359 & 0.612 & 1.301 & 1.251\\ 
\texttt{S2} & 1.219 & 0.726 & 1.483 & 0.754 & 6.025 & 0.387 & 6.506 & 0.444 & 1.875 & 1.697 \\ 
\texttt{S3} & 1.328 & 0.697 & 1.405 & 0.722 & 3.359 & 0.080 & 4.739 & 0.094 & 8.702 & 7.674 \\ 
\texttt{S4} & 1.436 & 0.663 & 1.343 & 0.695 & 3.142 & 0.022 & 4.243 & 0.024 & 30.48 & 29.02 \\ 
\texttt{L1} & 1.963 & 0.428 & 1.793 & 0.491 & 2.646 & 0.024 & 2.863 & 0.026 & 17.71 & 18.63 \\ 
\texttt{L2} & 1.576 & 0.613 & 1.297 & 0.646 & 2.987 & 0.129 & 3.576 & 0.144 & 4.759 & 4.477 \\ 
\texttt{L3} & 1.033 & 0.719 & 1.777 & 0.842 & 5.405 & 0.256 & 7.111 & 0.245 & 2.814 & 3.439 \\ 
\texttt{L4} & 1.560 & 0.828 & 1.839 & 0.873 & 6.537 & 0.210 & 7.886 & 0.213 & 3.943 & 4.100 \\
\hline  \vspace{2pt}                                   
\end{tabular}
\end{table*}
\section{Conclusions}
\label{sec:conclusions}
Following the work presented in Paper I, where we introduced a new formalism to investigate the dynamics of PWNe inside SNRs in the presence of a thick filamentary layer, that develops as the PWN expands inside the cold SNR ejecta, we have here extended that analysis to the following reverberation phase, in a three-structures model, where the filaments accumulate in a shell at the boundary between the PWN and SNR. We have implemented this formalism within a thin-shell Lagrangian code that follows the approach first introduced with the new \texttt{TIDE-L} code. \\
\\
The main findings of the present work are:
\begin{itemize}
    \item in the presence of a thick filamentary layer of cold ejecta, the duration of the free-expansion phase is shorter than in the standard thin-shell formalism, up to about 50\% for very energetic systems. This however does not translate into major differences in the typical size of PWNe at reverberation, which for a large portion of the population, is typically 0.65~$R_{\rm ch}$.
    \item the qualitative behavior during reverberation is unaltered: after reaching the reverse shock, the PWN expands a bit further up to a maximum radius, that differs at most $\sim $10\% with respect to the thin-shell case; then a compression sets in until a minimum radius is reached, which again does not differ substantially from the value in the thin-shell approximation; afterward the PWN undergoes a series of minor compressions and re-expansions.
    \item interestingly, the main difference with respect to the thin-shell case appears to be in the time it takes from the beginning of the reverberation to reach the first minimum (duration of the compression phase). In the presence of a thick layer of filaments, the duration of the first major compression is lengthened and the minimum is typically reached $\sim$ 50\% later in time. This smoother behavior is mainly a consequence of the difference in the way the mass accumulates at the PWN boundary.
    \item after having reached the first minimum, the PWN extension
    typically oscillates around values that are the same of the thin-shell case. Most of the difference between the two cases is limited just to the first compression.
    \item there is no appreciable difference in the measured compression factors, over a wide range of compressibilities. Given that this is the main quantity that affects the spectral evolution of middle-aged and old PWNe, potentially driving them to super-efficiency \citep{Torres_Lin+19a} and fast cooling, we do not expect the presence of a thick-layer of massive filaments to impact much on the long term spectral evolution of PWNe. 
\end{itemize}

We recall that this formalism, and the approach we have here introduced, while a step forward in understanding and modeling the role played by  a filamentary layer in the dynamics and evolution of PWNe, is still limited to a pure 1D-radial evolution. In this sense, it cannot answer questions related to the multi-dimensional stability of the compression phase. Simulations and observations \citep{Blondin_Chevalier+01a,Kolb_Blondin+17a,Ma_Ng+10a,Meyer_Torres25a} clearly indicate that mixing of the PWN and SNR is likely to be important during reverberation. Handling such complex physics is possible only using multi-dimensional numerical codes for fluid dynamics. 
However, from a numerical perspective, following the entire evolution of a PWN to very late times (from a few years to several tens of kyr), in the appropriate regimes (combining a relativistic PWN with a non relativistic SNR), with the correct energetics (including the PSR spin-down), and with sufficient resolution to resolve the formation of a filamentary layer, not to mention the inclusion of magnetic field, 
is still not feasible from a numerical point of view, being too demanding in terms of time and resources.
How much and how large the role of any possible instability would be, and their impact on the spectral properties of PWNe, still remain to be assessed.
\begin{acknowledgements}
The authors thank D.~F. Torres for fruitful comments and suggestion on this work. N.B. and B.O. acknowledge partial support by the INAF Mini-grant HYPNOTIC87A: "Hidden Young Pulsar Nebula Occupying The Inner Core of 87A" and by the European Union – NextGenerationEU RRF M4C2 1.1 grant PRIN-MUR 2022TJW4EJ.
\end{acknowledgements}

\bibliographystyle{aa}
\bibliography{article_2}

\end{document}